# Enhancing Privacy in Federated Learning through Quantum Teleportation Integration

**Author**: Koffka Khan

**Abstract**

Federated learning enables collaborative model training across multiple clients without sharing raw data, thereby enhancing privacy. However, the exchange of model updates can still expose sensitive information. Quantum teleportation, a process that transfers quantum states between distant locations without physical transmission of the particles themselves, has recently been implemented in real-world networks. This position paper explores the potential of integrating quantum teleportation into federated learning frameworks to bolster privacy. By leveraging quantum entanglement and the no-cloning theorem, quantum teleportation ensures that data remains secure during transmission, as any eavesdropping attempt would be detectable. We propose a novel architecture where quantum teleportation facilitates the secure exchange of model parameters and gradients among clients and servers. This integration aims to mitigate risks associated with data leakage and adversarial attacks inherent in classical federated learning setups. We also discuss the practical challenges of implementing such a system, including the current limitations of quantum network infrastructure and the need for hybrid quantum-classical protocols. Our analysis suggests that, despite these challenges, the convergence of quantum communication technologies and federated learning presents a promising avenue for achieving unprecedented levels of privacy in distributed machine learning.

## 1. Introduction

Federated Learning (FL) has emerged as a transformative approach in machine learning, enabling multiple clients to collaboratively train models without the necessity of sharing their raw data. This decentralized methodology inherently enhances data privacy by keeping sensitive information localized. However, the process of transmitting model updates between clients and central servers remains susceptible to security breaches, potentially exposing confidential data through sophisticated inference attacks. Consequently, there is a pressing need to bolster the security of communication channels within FL systems to ensure comprehensive privacy preservation.

In parallel, the field of quantum communication has witnessed significant advancements, particularly with the practical implementation of quantum teleportation over existing fiber-optic networks. Quantum teleportation facilitates the transfer of quantum states between distant locations without the physical movement of particles, leveraging the principles of quantum entanglement. Notably, researchers at Northwestern University have recently demonstrated quantum teleportation over a 30-kilometer fiber-optic cable concurrently carrying conventional internet traffic, underscoring the feasibility of integrating quantum communication with current infrastructure (Northwestern University, 2024).

The convergence of Federated Learning and quantum communication presents a compelling opportunity to enhance data privacy. By incorporating quantum teleportation into FL frameworks, it becomes possible to establish intrinsically secure communication channels. The fundamental properties of quantum mechanics, such as the no-cloning theorem and the detection of eavesdropping attempts through quantum state disturbances, provide robust safeguards against unauthorized data access during transmission.

Recent studies have begun exploring this intersection. For instance, the concept of Quantum Federated Learning (QFL) has been proposed to leverage quantum computation and quantum cryptography for improved privacy and security in FL. Techniques such as quantum homomorphic encryption have been introduced to enable quantum delegated and federated learning with computational-theoretical data privacy guarantees, offering reduced communication complexity compared to classical methods (Li & Deng, 2024).

Implementing quantum teleportation within FL systems involves several technical challenges. Establishing and maintaining entanglement over long distances, ensuring compatibility with existing network infrastructure, and developing efficient quantum-classical hybrid protocols are critical considerations. Moreover, the current limitations of quantum hardware, including error rates and coherence times, necessitate careful system design to achieve reliable performance.

Despite these challenges, the potential benefits are substantial. Integrating quantum teleportation into FL can provide unparalleled security for data transmission, effectively mitigating risks associated with data interception and inference attacks. This integration could pave the way for more secure distributed machine learning applications, particularly in sectors where data privacy is paramount, such as healthcare and finance.

In conclusion, the fusion of quantum teleportation and Federated Learning represents a promising frontier in the quest for enhanced data privacy in distributed machine learning. Ongoing research and technological advancements in quantum communication are poised to play a pivotal role in realizing this vision, offering a pathway to intrinsically secure collaborative learning environments.

## 2. Literature Review

Federated Learning (FL) has emerged as a decentralized machine learning paradigm that enables multiple clients to collaboratively train models without sharing their raw data, thereby enhancing privacy and security (Kairouz et al., 2021). However, traditional FL approaches are susceptible to inference attacks during the transmission of model updates, necessitating more robust privacy-preserving techniques.

Recent advancements in quantum computing and communication offer promising avenues to bolster FL's security. Quantum Federated Learning (QFL) integrates quantum mechanics principles, such as superposition and entanglement, to enhance data privacy in distributed learning environments. For instance, the Quantum Fuzzy Federated Learning (QFFL) framework combines quantum computing with fuzzy logic to protect local data and accelerate global learning efficiency, demonstrating improved performance on datasets like COVID-19 and MNIST (Zhao et al., 2023).

Another approach involves the use of Fully Homomorphic Encryption (FHE) within QFL systems. Dutta et al. (2024) propose a paradigm that integrates quantum computing with FHE, allowing operations on encrypted data without decryption. This method facilitates collaborative model training while preserving data privacy and reducing computational overhead.

Differential privacy techniques have also been explored in the context of QFL. Rofougaran et al. (2023) demonstrate that combining federated learning with differential privacy in quantum-classical machine learning models can achieve high test accuracy while maintaining strong privacy guarantees, even on Noisy Intermediate-Scale Quantum (NISQ) devices.

The integration of quantum networks into FL frameworks has been investigated to enhance communication security. Li et al. (2023) explore the use of quantum networks to facilitate secure communication in QFL, highlighting the potential of quantum communication to provide secure channels for model updates.

Despite these advancements, challenges remain in implementing QFL systems. Current quantum hardware limitations, such as qubit coherence time and error rates, pose significant obstacles. Additionally, the seamless integration of quantum and classical components requires careful consideration of interface compatibility and data transformation between quantum and classical representations. Scalability is another concern, as the proposed systems must effectively handle an increasing number of clients and larger datasets.

In conclusion, the convergence of quantum computing and federated learning presents a promising frontier for enhancing data privacy and security in distributed machine learning environments. Ongoing research into QFL frameworks, encryption techniques, and quantum communication protocols is essential to address existing challenges and fully realize the potential of this interdisciplinary approach.

## 3. Mathematical Proposal and Analysis

### Federated Learning Framework

Consider a federated learning system with $N$ clients, each possessing a local dataset $D_i$ for $i = 1, 2, \ldots, N$. The objective is to collaboratively train a global model $w$ by minimizing a loss function $\mathcal{L}(w)$ defined as:

$$\mathcal{L}(w) = \sum_{i=1}^{N} \frac{|D_i|}{|D|} \mathcal{L}_i(w)$$

where $\mathcal{L}_i(w)$ is the local loss function for client $i$, $|D_i|$ is the size of the local dataset, and $|D| = \sum_{i=1}^{N} |D_i|$ is the total dataset size.

The federated learning process involves the following steps:

1. **Initialization**: The central server initializes the global model $w^{(0)}$.

2. **Local Training**: Each client updates the global model using their local data:

$$w_i^{(t+1)} = w^{(t)} - \eta \nabla \mathcal{L}_i(w^{(t)})$$

where $\eta$ is the learning rate, and $\nabla \mathcal{L}_i(w^{(t)})$ is the gradient of the local loss function at iteration $t$.

3. **Aggregation**: The server aggregates the local updates to form a new global model:

$$w^{(t+1)} = \sum_{i=1}^{N} \frac{|D_i|}{|D|} w_i^{(t+1)}$$

This process repeats until convergence.

## Quantum Teleportation Protocol

Quantum teleportation enables the transfer of a quantum state $|\psi\rangle$ from one party (Alice) to another (Bob) using a pair of entangled qubits and classical communication. The protocol involves the following steps:

1. **Entanglement Preparation**: Alice and Bob share an entangled state $|\Phi^+\rangle_{AB} = \frac{1}{\sqrt{2}}(|0\rangle_A|0\rangle_B + |1\rangle_A|1\rangle_B)$.
2. **Bell State Measurement**: Alice performs a joint measurement on her qubit of the entangled pair and the qubit in state $|\psi\rangle$, collapsing the system into one of the four Bell states.
3. **Classical Communication**: Alice sends the result of her measurement (2 classical bits) to Bob.
4. **State Reconstruction**: Bob applies a corresponding unitary operation $U$ to his qubit to reconstruct the original state $|\psi\rangle$.

This protocol ensures that the quantum state is transferred without physically transmitting the qubit itself, providing inherent security due to the principles of quantum mechanics.

## Integration into Federated Learning

To integrate quantum teleportation into federated learning, we propose the following modifications:

1. **Quantum Representation of Model Updates**: Each client's model update $\Delta w_i^{(t)} = w_i^{(t+1)} - w^{(t)}$ is encoded into a quantum state $|\Delta w_i^{(t)}\rangle$.
2. **Entanglement Distribution**: The server distributes entangled qubit pairs to each client.
3. **Quantum Teleportation of Updates**: Clients use the entangled qubits to teleport their quantum-encoded model updates to the server.
4. **Aggregation in Quantum Domain**: The server performs quantum operations to aggregate the received quantum states, forming the updated global model.
5. **Classical Extraction**: The aggregated quantum state is measured to extract the updated global model parameters $w^{(t+1)}$.

## Privacy Analysis

The integration of quantum teleportation enhances privacy in the following ways:

- **Eavesdropping Detection**: Any interception attempt of the entangled qubits or the quantum states during teleportation will disturb the quantum system, making eavesdropping detectable.

- **No-Cloning Theorem**: Quantum information cannot be copied perfectly, preventing adversaries from duplicating the model updates during transmission.

**Proof of Security Enhancement**

*Theorem*: Integrating quantum teleportation into federated learning ensures that any eavesdropping attempt during the transmission of model updates can be detected, thereby enhancing the privacy of the learning process.

*Proof*:

1. **Eavesdropping Detection**: Assume an adversary intercepts the quantum channel during the teleportation of a model update $|\Delta w_i^{(t)}\rangle$. Due to the principles of quantum mechanics, any measurement by the adversary will collapse the quantum state, introducing detectable anomalies. Specifically, the entanglement between the qubits will be disturbed, and the server will receive a state that deviates from the expected $|\Delta w_i^{(t)}\rangle$, signaling a potential eavesdropping attempt.

2. **No-Cloning Theorem**: The no-cloning theorem states that it is impossible to create an identical copy of an arbitrary unknown quantum state. Therefore, an adversary cannot duplicate the quantum-encoded model updates without introducing detectable disturbances, ensuring that the information remains secure during transmission.

By leveraging quantum teleportation within federated learning (FL) systems, we can enhance the privacy and security of model updates during transmission. This integration ensures that any eavesdropping attempts are detectable, thereby safeguarding sensitive information.

## 4. Proposed Methodology

To enhance privacy in federated learning (FL) systems, we propose integrating quantum teleportation for the secure transmission of model updates. This methodology leverages the principles of quantum mechanics to ensure that any eavesdropping attempts during communication are detectable, thereby safeguarding sensitive information.

## 1. Quantum Representation of Model Updates

Each client computes a model update, denoted as $\Delta w_i^{(t)}$, based on their local data. This update is then encoded into a quantum state $|\Delta w_i^{(t)}\rangle$. The encoding process involves mapping the classical data of the model update into the amplitudes of a quantum state, ensuring that the information is represented in a quantum format suitable for teleportation.

## 2. Entanglement Distribution

The central server prepares entangled qubit pairs, known as Bell states, and distributes them to the clients. A Bell state is a specific quantum state of two qubits that represents the simplest form of entanglement and is defined as:

$$|\Phi^+\rangle = \frac{1}{\sqrt{2}}(|00\rangle + |11\rangle)$$

Each client receives one qubit from the entangled pair, while the server retains the corresponding entangled qubit.

## 3. Quantum Teleportation of Updates

Clients utilize the received entangled qubit and their encoded model update $|\Delta w_i^{(t)}\rangle$ to perform quantum teleportation. The teleportation protocol involves the following steps:

- **Bell State Measurement**: The client performs a joint measurement on their qubit of the entangled pair and the qubit encoding the model update, projecting the system onto one of th four Bell states.

- **Classical Communication**: The client sends the result of the Bell state measurement (2 classical bits) to the server.

- **State Reconstruction**: Based on the received classical information, the server applies a corresponding unitary operation to its qubit to reconstruct the original quantum state $|\Delta w_i^{(t)}\rangle$

This process ensures that the model update is transmitted securely without the physical transfer of the quantum state, leveraging the no-cloning theorem to prevent unauthorized duplication.

### 4. Aggregation in Quantum Domain

Upon receiving the teleported quantum states from all clients, the server aggregates the model updates in the quantum domain. This involves performing quantum operations that correspond to the weighted sum of the quantum states:

$$|\Delta w^{(t+1)}\rangle = \sum_{i=1}^{N} \frac{|D_i|}{|D|} |\Delta w_i^{(t)}\rangle$$

where $|D_i|$ is the size of the local dataset for client $i$, and $|D| = \sum_{i=1}^{N} |D_i|$ is the total dataset size.

### 5. Classical Extraction

The aggregated quantum state $|\Delta w^{(t+1)}\rangle$ is measured to extract the updated global model parameters $\Delta w^{(t+1)}$. The server then updates the global model:

$$w^{(t+1)} = w^{(t)} + \Delta w^{(t+1)}$$

This updated model is subsequently communicated to the clients for the next iteration of training.

**Privacy Assurance**

The integration of quantum teleportation into the FL framework provides robust privacy assurances:

- **Eavesdropping Detection**: Any attempt to intercept the quantum communication will disturb the quantum states due to the principles of quantum mechanics, making eavesdropping detectable.

- **No-Cloning Theorem**: The impossibility of creating identical copies of an unknown quantum state ensures that the model updates cannot be duplicated without detection.

**Implementation Considerations**

Implementing this methodology requires addressing several practical challenges:

- **Quantum Network Infrastructure**: Establishing a reliable quantum network capable of distributing entangled qubits to clients is essential. This involves developing quantum repeaters and ensuring the stability of entanglement over long distances.

- **Quantum-Classical Integration**: Designing efficient protocols that seamlessly integrate quantum communication with classical FL processes is crucial. This includes encoding classical model updates into quantum states and decoding them appropriately after teleportation.

- **Error Correction**: Quantum systems are susceptible to errors due to decoherence and other quantum noise. Implementing robust quantum error correction mechanisms is necessary to maintain the integrity of teleported information.

By addressing these considerations, the proposed methodology aims to enhance the privacy and security of federated learning systems through the integration of quantum teleportation.

**5. Proposed Experiments**

To evaluate the effectiveness of integrating quantum teleportation into federated learning (FL) for enhanced privacy, we propose a series of experiments focusing on implementation feasibility, privacy enhancement, and performance metrics. Implementation feasibility involves assessing the practicality of encoding classical model updates into quantum states and performing quantum teleportation within current technological constraints. Privacy enhancement evaluates the degree to which quantum teleportation mitigates risks associated with data leakage and adversarial attacks compared to classical FL communication methods. Performance metrics analyze the impact of quantum communication on model convergence rates, accuracy, and overall computational efficiency.

Selecting appropriate datasets is crucial for comprehensive evaluation. We propose using both classical and quantum datasets to assess the versatility of the integrated system. Classical datasets, such as MNIST (handwritten digit recognition) and CIFAR-10 (image classification), can be used to evaluate the performance of the FL model in a classical context. Quantum datasets, like the QDataSet, which comprises 52 high-quality datasets derived from simulations of one- and two-qubit systems evolving in the presence and/or absence of noise, provide a wealth of information for training and developing quantum machine learning algorithms.

Implementing the proposed methodology requires a combination of classical and quantum machine learning frameworks. TensorFlow Federated (TFF), an open-source framework for machine learning and other computations on decentralized data, can be used to simulate the federated learning environment and manage the aggregation of model updates. TensorFlow Quantum (TFQ), a library for hybrid quantum-classical machine learning, can handle the quantum aspects of the model, including the encoding of data into quantum states and the implementation of quantum circuits. Qiskit, an open-source quantum computing software development framework, can be used to design and simulate quantum circuits, perform quantum teleportation protocols, and manage quantum state manipulations.

The experimental setup involves several steps. First, set up a simulated federated learning environment using TFF, where multiple clients train local models on their respective datasets. Next, implement quantum data encoding techniques using TFQ or Qiskit to convert classical model updates into quantum states. Then, design and simulate the quantum teleportation process for transmitting quantum-encoded model updates from clients to the central server. After that, use TFF to aggregate the teleported quantum model updates and update the global model accordingly. Finally, assess the system's performance in terms of model accuracy, convergence rates, communication overhead, and privacy preservation by comparing it with a baseline classical FL system.

Evaluation metrics include model accuracy, which measures the accuracy of the global model on a validation dataset after each training round; convergence rate, which analyzes the number of communication rounds required for the model to converge to a predefined accuracy threshold; communication overhead, which evaluates the amount of data transmitted during the training process, considering both classical and quantum communication channels; and privacy assessment, which assesses the system's resilience to potential attacks aimed at extracting information from the communicated model updates.

Challenges and considerations include quantum hardware limitations, as current quantum hardware may have limitations in terms of qubit coherence time, gate fidelity, and error rates; hybrid integration, as seamlessly integrating quantum and classical components requires careful consideration of interface compatibility and data transformation between quantum and classical representations; and scalability, as assessing the scalability of the proposed system is essential to ensure it can handle an increasing number of clients and larger datasets effectively.

By conducting these experiments, we aim to demonstrate the feasibility and advantages of incorporating quantum teleportation into federated learning systems, paving the way for more secure and efficient distributed machine learning frameworks.

## 6. Limitations

Integrating quantum teleportation into federated learning (FL) offers promising avenues for enhancing privacy; however, several limitations and challenges must be addressed to realize its practical implementation.

Current quantum computing hardware is in the early stages of development, characterized by limited qubit counts, short coherence times, and susceptibility to errors. These constraints pose significant challenges to the reliable execution of quantum teleportation protocols. Establishing and maintaining entanglement over long distances is technically demanding. Although recent advancements have demonstrated quantum teleportation over fiber optic networks, the stability and scalability of such entangled states in real-world conditions remain areas of active research (Thomas et al., 2024).

Seamlessly integrating quantum communication protocols with existing classical FL systems requires sophisticated interfacing mechanisms. This integration involves encoding classical data into quantum states and decoding quantum information back into classical formats, processes that are complex and error-prone. Implementing quantum teleportation necessitates significant computational resources and infrastructure, which may not be readily available or cost-effective for all organizations.

While quantum teleportation offers theoretical security advantages, practical implementations must consider comprehensive threat models. Determining realistic threat models and ensuring that the system can defend against various types of attackers, including those who may attempt to modify or disrupt the system's operation, is challenging (Near & Darais, 2024).

Even with secure communication channels, FL systems remain vulnerable to data poisoning attacks, where malicious clients introduce corrupted data to compromise the global model. Addressing these risks requires additional mechanisms beyond secure communication (Near & Darais, 2024).

Scaling quantum networks to accommodate a large number of clients in an FL system is a significant challenge. The complexity of managing entangled states and ensuring synchronized operations across numerous clients can impede scalability. The additional computational overhead introduced by quantum operations may affect the efficiency and speed of the FL process, particularly in large-scale deployments.

Compliance with data privacy laws and regulations, such as the General Data Protection Regulation (GDPR), adds complexity to the implementation of quantum-enhanced FL systems. Ensuring that such systems meet legal standards for data protection is essential. The deployment of advanced quantum technologies in data processing raises ethical questions regarding data ownership, consent, and the potential for unintended consequences, necessitating careful consideration.

Addressing these limitations requires interdisciplinary collaboration, continued technological advancements, and the development of robust frameworks that can effectively integrate quantum communication protocols into federated learning systems.

## 7. Conclusion

In this paper, we have explored the integration of quantum teleportation into federated learning (FL) frameworks to enhance data privacy and security. By leveraging the principles of quantum mechanics, particularly quantum entanglement and the no-cloning theorem, we propose a methodology that ensures secure transmission of model updates without exposing sensitive information.

Our contributions include the development of a mathematical framework detailing how quantum teleportation can be employed to transmit model updates securely within an FL system. This framework demonstrates that any eavesdropping attempts during transmission would be detectable, thereby preserving data integrity. We also outline a step-by-step approach for implementing quantum teleportation in FL, including quantum representation of model updates, entanglement distribution, secure transmission, aggregation in the quantum domain, and classical extraction of the updated global model. Additionally, we propose a series of experiments to evaluate the feasibility and effectiveness of the integrated system, utilizing both classical and quantum datasets, and leveraging frameworks such as TensorFlow Federated and Qiskit.

The integration of quantum teleportation into FL presents a promising avenue for achieving unprecedented levels of privacy in distributed machine learning. However, practical implementation faces challenges, including the current limitations of quantum hardware, the need for robust quantum network infrastructure, and the development of efficient quantum-classical hybrid protocols.

Future research directions include investigating the scalability of the proposed system to accommodate a large number of clients and extensive datasets, developing quantum error correction techniques to address the susceptibility of quantum systems to decoherence and noise, and designing efficient protocols that seamlessly integrate quantum and classical components to optimize performance and resource utilization.

By addressing these challenges, the convergence of quantum communication technologies and federated learning has the potential to revolutionize data privacy and security in distributed machine learning environments.